\documentclass[twocolumn,superscriptaddress,showpacs,%
preprintnumbers,amsmath,amssymb]{revtex4}
\usepackage{amsmath}
\usepackage{graphicx}
\usepackage{dcolumn}
\newcommand{\p}{\partial}
\begin{document}

\title{Structure and correlations in two-dimensional \\Coulomb confined classical artificial atoms}
\author{W.~P.~Ferreira}
\email{paiva@fisica.ufc.br}
\affiliation{Departamento de F\'isica,
Universidade Federal do
Cear\'a, Caixa Postal 6030, Campus do Pici, 60455-760 Fortaleza, Cear\'a, Brazil}%
\author{A.~Matulis}
\email{amatulis@takas.lt}
\altaffiliation[permanent address:\ ]
{Institute of Semiconductor Physics, Gostauto 11, 2600 Vilnius,
Lithuania}
\affiliation{Departement Natuurkunde, Universiteit
Antwerpen (UIA), Universiteitsplein 1, B-2610 Antwerpen, Belgium}
\author{G.~A.~Farias}
\email{gil@fisica.ufc.br} \affiliation{Departamento de F\'isica,
Universidade Federal do
Cear\'a, Caixa Postal 6030, Campus do Pici, 60455-760 Fortaleza, Cear\'a, Brazil}%
\author{F.~M.~Peeters}
\email{peeters@uia.ua.ac.be}
\affiliation{Departement Natuurkunde,
Universiteit Antwerpen (UIA), Universiteitsplein 1, B-2610
Antwerpen, Belgium}

\date{October 29, 2002}

\begin{abstract}
The ordering of $N$ equally charged particles ($-e$) moving in two
dimensions and confined by a Coulomb potential, resulting from a
displaced positive charge $Ze$ is discussed. This is a classical
model system for atoms. We obtain the configurations of the
charged particles which, depending on the value of $N$ and $Z$,
may result in ring structures, hexagonal-type configurations and
for $N/Z \approx 1$ in an inner structure of particles which is
separated by an outer ring of particles. For $N/Z << 1$ the
Hamiltonian of the parabolic confinement case is recovered. For
$N/Z \approx 1$ the configurations are very different from those
found in the case of a parabolic confinement potential. A
hydrodynamic analysis is presented in order to highlight the
correlations effects.
\end{abstract}


\maketitle

\section{Introduction}

Quantum dots, or \textit{artificial atoms}, have been a subject of
intense theoretical and experimental studies in recent years
\cite{jacak98} due to the occurrence of numerous interesting
effects caused by electron correlation, such as e.~g.~Wigner
crystallization, overcharging, and nontrivial behavior in a
magnetic field. These electron correlations appear most clearly
when the electron interaction and the confinement potential
dominates over the kinetic energy of the system. This can be
realized in e.g.~quantum dots which have much smaller electron
density as compared to real atoms.

Correlation effects show up in an even more pronounced way in
classical systems where the kinetic energy is zero in the absence
of thermal fluctuations. Two-dimensional ($2D$) classical dots
confined by a parabolic potential were studied earlier and a table
of Mendeleyev for such "artificial" atoms was constructed
\cite{bedanov94}. The classical system which is more closely
related to real atoms was studied in Ref. \cite{gil96} where, as
function of the strength of the confinement potential, surprising
rich physics was observed like structural transitions, spontaneous
symmetry breaking, and unbinding of particles, which is absent in
parabolic confined dots. The confinement potential was of the
Coulomb type, but in order to prevent the collapse of all
electrons onto the "nucleus" (which we call 'impurity' in this
case) the positive charge ($Ze$) was displaced a certain distance
from the $2D$ plane where the electrons are moving in (see
Fig.~\ref{fig:figModel}). Note that our system is related to the
"superatom" system introduced by Watanabe and Inoshita
\cite{watanabe86}. The superatom is a spherical modulation-doped
heterojunction. In particular it is a quasi-atomic system which
consists of a spherical donor-doped core and a surrounding
impurity-free matrix with a larger electron affinity. The quantum
mechanical electron structure of this system was studied in
\cite{inshita88} and it was found that due to the absence of the
$1/r$ singularity in the potential the ordering of the energy
levels is dominated by the no-radial-node states, in contrast to
real atoms where $s$ and $p$ states are dominant.

In the present paper we present a systematic study of the system
of Ref. \cite{gil96} as a function of the number of particles
($N$). In contrast to Ref. \cite{gil96} we will not vary the
strength of the confinement potential, as this was already
presented in our earlier work \cite{gil96} where we limited the
numerical results to the case $Z=N=4$, but we will vary Z and N.
Furthermore, we will
compare our numerical results with the results of a hydrodynamic,
i.e. continuum, approach. In the latter case the electron density
is taken as a fluid, i.e. there are no charge quanta. This is in
contrast to our numerical simulation where the electrons are point
particles with a fixed quantized charge value. The fact that
charge is now distributed in packages of $q=-e$ will introduce
important correlation effects which is the central theme of the
present work.

Note also that the present classical study can serve as a zeroth
order approach for more demanding quantum mechanical calculations.
Recently, such classical calculations were used \cite{yannou} as a
starting point in order to construct better quantum wavefunctions,
at least in the strong magnetic field limit.

Besides the above mentioned analogy with real and artificial atoms
which are inherently quantum mechanical there exist other
experimental realized systems which behave purely classical and
for which our study is relevant. Examples are charged colloidal
suspensions where it was found recently that correlation effects
between the counter-ions can result into an overscreening and
attraction between like charged colloids \cite{colloid}. Our
system is a simplified 2D model for those colloidal systems.

The system under study can be realized experimentally in the
system of electrons above liquid helium \cite{andrei} by putting a
positive localized charge in the substrate which supports the
liquid helium. The equivalent quantum mechanical system can be
realized using low dimensional semiconductor structures with
impurities, also called remote impurities, which are displaced a
distance from a quantum well \cite{marmorkos}. In both cases it will
be rather difficult to increase the number of positive charge $Z$
beyond a few units.

Another possible realization of our system is by bringing an $AFM$
(atomic force microscope) tip close to a $2D$ electron gas. When
this tip is charged positively it will induce a confinement
potential very similar to the one studied in the present paper.
The advantage of this approach is that the charge $Ze$ on the tip
can be varied continuously by increasing the voltage on the tip.

This paper is organized as follows. In section II we describe the
mathematical model and our numerical approach to obtain the
configurations. The results of our numerical simulations are given
in section III. A hydrodynamic approach which neglects the
correlation effects is presented in sections IV. The numerical
simulation results are compared with the results of the
hydrodynamic approximation in section V which clearly brings about
the importance of the charge correlation. Our conclusions are
presented in section VI.

\section{The Model}
\label{sec:model}
\begin{figure} \centering
\vspace{5mm}
\begin{center}
\includegraphics[scale=0.50]{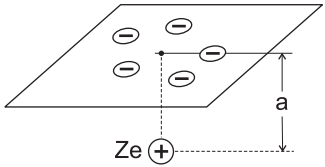}
\caption{Schematic view of the system.}
\label{fig:figModel}
\end{center}
\end{figure}
We study a system with $N$ negatively charged particles ($-e$),
which we call here and further electrons, interacting through a repulsive
Coulomb potential and moving in the $xy$-plane. The particles are
kept together through a fixed positive charge ($Ze$) located at a
distance $a$ from the plane the particles are moving in (see
Fig.~\ref{fig:figModel}). The total energy of this system is given
by the Hamiltonian
\begin{equation}\label{hamiltonian}
\displaystyle{
  H = -\frac{Ze^{2}}{\epsilon}\sum_{i=1}^{N}\frac{1}{\sqrt{r_i^2+a^2}}
  + \frac{e^{2}}{\epsilon}\sum_{i>j=1}^{N}
  \frac{1}{\vert {\bf r}_{i} - {\bf r}_{j}\vert}.
}
\end{equation}
Here the symbol $\epsilon$ stands for the dielectric constant,
and ${\bf r}=\{x,y\}$ is the two component position vector of the 2D
(two dimensional) electron.
For convenience, we express the electron energy in units of
$E_{0}=e^{2}/{\epsilon}a$ and all the distances in units of $a$.
This allows us to rewrite Eq.~(\ref{hamiltonian}) in the
following dimensionless form:
\begin{equation}\label{reducedhamilt}
\displaystyle{
  H = -\sum_{i=1}^{N}\frac{Z}{\sqrt{r_i^2+1}} +
  \sum_{i>j=1}^{N}\frac{1}{\vert {\bf r}_{i} - {\bf r}_{j}\vert}.
}
\end{equation}

The ground state configurations of the two-dimensional system were
obtained using the standard Metropolis algorithm
\cite{metropolis}. The electrons are initially put in random
positions within some circle and allowed to reach a steady state
configuration after a number of simulation steps in the order of
$10^{5}$. To check if the obtained configuration is stable, we
calculated the frequencies of the normal modes of the system using
the Householder diagonalization technique \cite{schweg}. The
configuration was taken as final when all frequencies of the
normal modes are positive and the energy did not decrease further.
The meta-stable states were avoided introducing a small
temperature $T=10^{-4}$, which was negligible and does not
influence the simulation accuracy.

\section{Stable configurations}

\begin{figure}
\begin{center}
\includegraphics[scale=0.30,angle=-90]{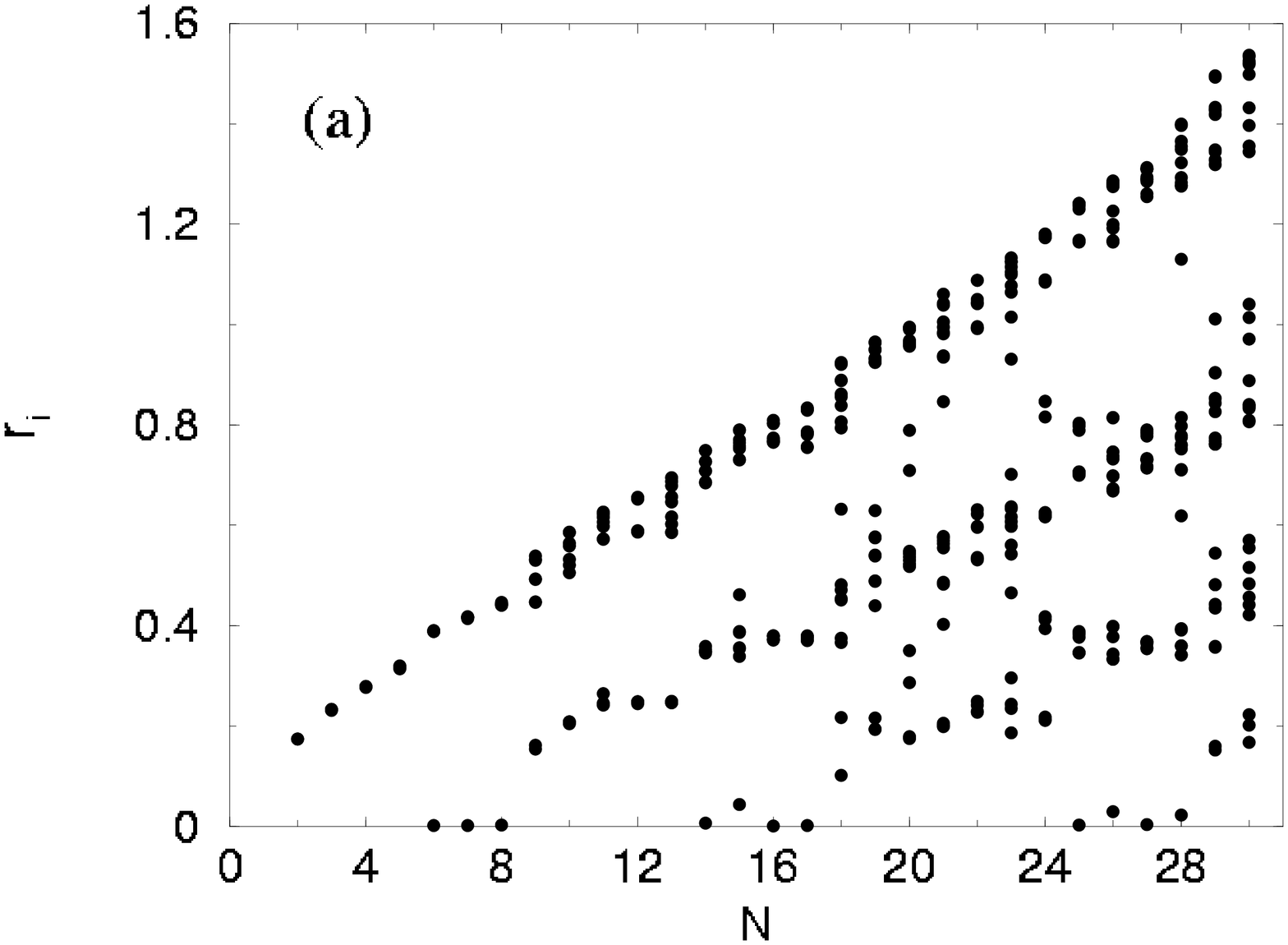}
\end{center}
\vspace{6mm}
\begin{center}
\includegraphics[scale=0.30,angle=-90]{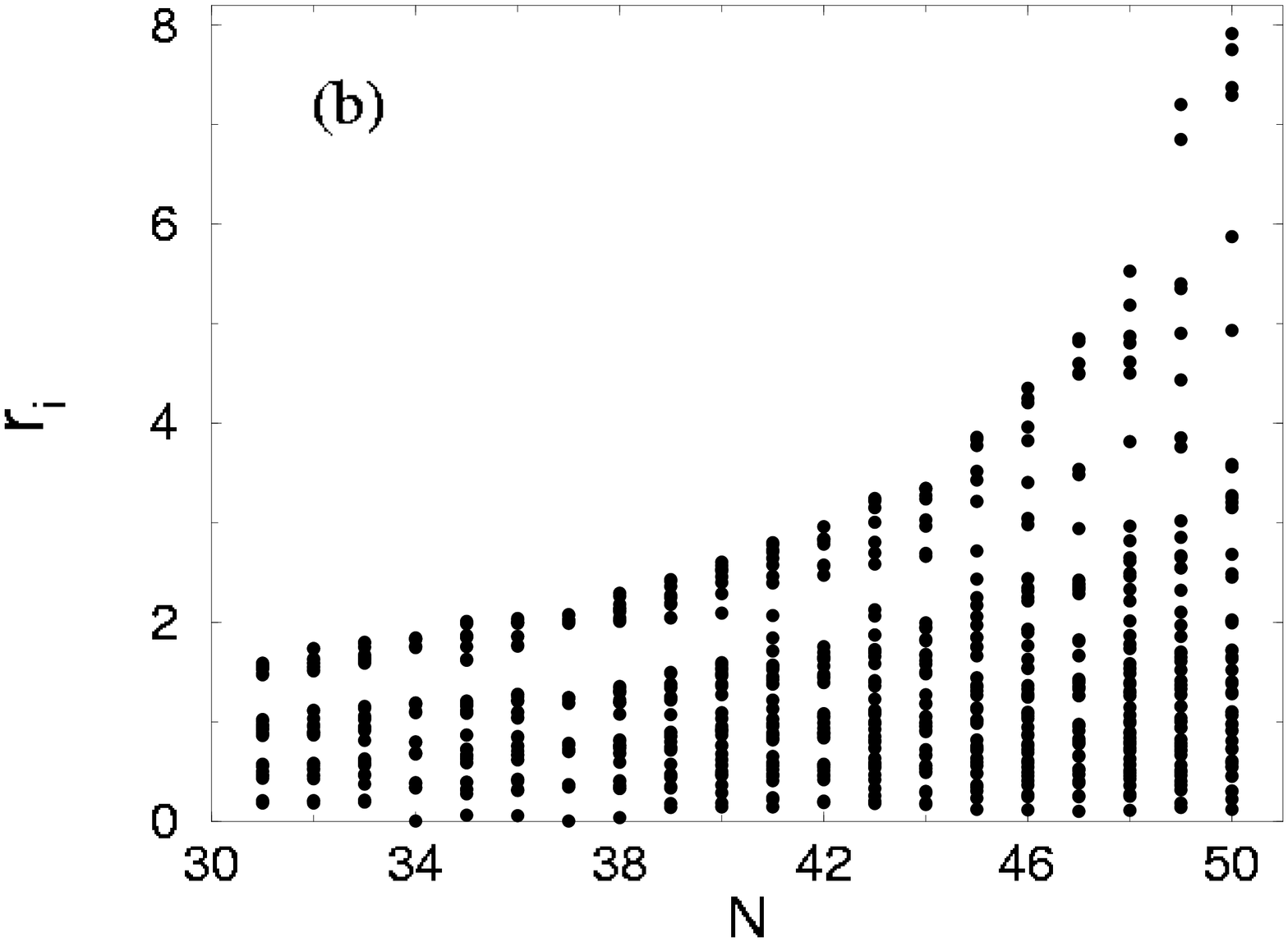}
\vspace{3mm} \caption{The radial position of the particles, $r_i$,
as a function of the number of electrons in the dot, $N$, for a
system with confinement positive charge $Z=50$. The number of
electrons is varied from $N=2$ up to $N=50$. } \label{fig:fig2}
\end{center}
\end{figure}

\begin{figure} \centering
\vspace{5mm}
\includegraphics[scale=0.70]{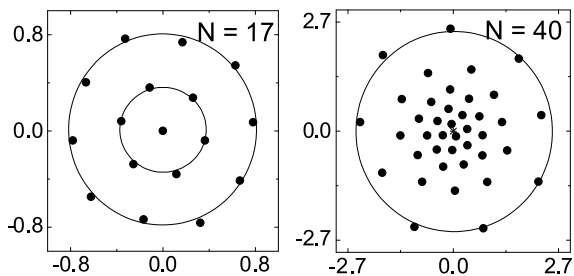}
\caption{Configurations for two values of the number of particles
$N$ for a fixed Coulomb center with $Z=50$.} \label{fig:fig3}
\end{figure}

As an example of our results we present in Fig.~\ref{fig:fig2} the
radial distribution of the electrons for a fixed positive charge
$Z=50$ for the impurity as function of the numbers of electrons
$N$ in the dot.
Clearly, two different electron distribution types can be
distinguished. Namely, for small number of electrons $N$ a ring
structure arrangement for the electrons is observed which is
similar to the one for a parabolic dot (compare with the
configurations given in \cite{bedanov94}), for large $N$ the outer
electrons can form a ring which is clearly separated from the
other electrons in the dot. This is more clearly seen in
Fig.~\ref{fig:fig3} where examples of two different configurations
(for small and large $N$) are presented.
Note that in the case of large $N$ the core electrons
are arranged in nearly a triangular lattice which is characteristic
for an infinite electron system.

The very different type of configurations for small $N$ and large
$N$ is a consequence of the fact that the screening of the Coulomb
center is a much more complicated problem as compared with the
parabolic dot case. Now its behavior is controlled by two
parameters, namely, the number of electrons $N$, which
characterizes the discreteness of the charges taking participation
in the screening, and the positive charge $Z$, which represents
the strength of the confinement Coulomb potential.

In the case of a relatively small number of electrons ($n=N/Z\ll 1$)
the confinement potential is strong, and the electrons are located close
to the origin where the confinement potential can be replaced by
the following approximate one:
\begin{equation}\label{vapp}
  -\frac{Z}{\sqrt{r^2+1}} \approx V_{appr}(r)
  = -Z + \frac{Z}{2}r^2.
\end{equation}
Now substituting it into Hamiltonian (\ref{reducedhamilt}),
and scaling the variables
\begin{equation}\label{scale}
  H \to (Z/2)^{1/3}H, \quad {\bf r}\to (2/Z)^{1/3}{\bf r},
\end{equation}
one obtains the Hamiltonian
\begin{equation}\label{hambp}
  H_{appr} = -\left(2Z^2\right)^{1/3}N + \sum_{i=1}^N r_i^2
  + \sum_{i>j=1}^N\frac{1}{|{\bf r}_i- {\bf r}_j|},
\end{equation}
which, up to a non-essential energy shift, coincides with the
Hamiltonian of a parabolic dot as considered in reference
\cite{bedanov94}. In the above $Z=50$ case we have the same
parabolic dot like configurations up to $N=9$, while for $N\geq
10$ new configurations as (3,7), (4,7) and so on appear. For
larger $Z$ the similarity between parabolic dot and Coulomb center
screening persists up to larger $N$ values.

\begin{figure}
\begin{center}
\includegraphics[scale=0.6]{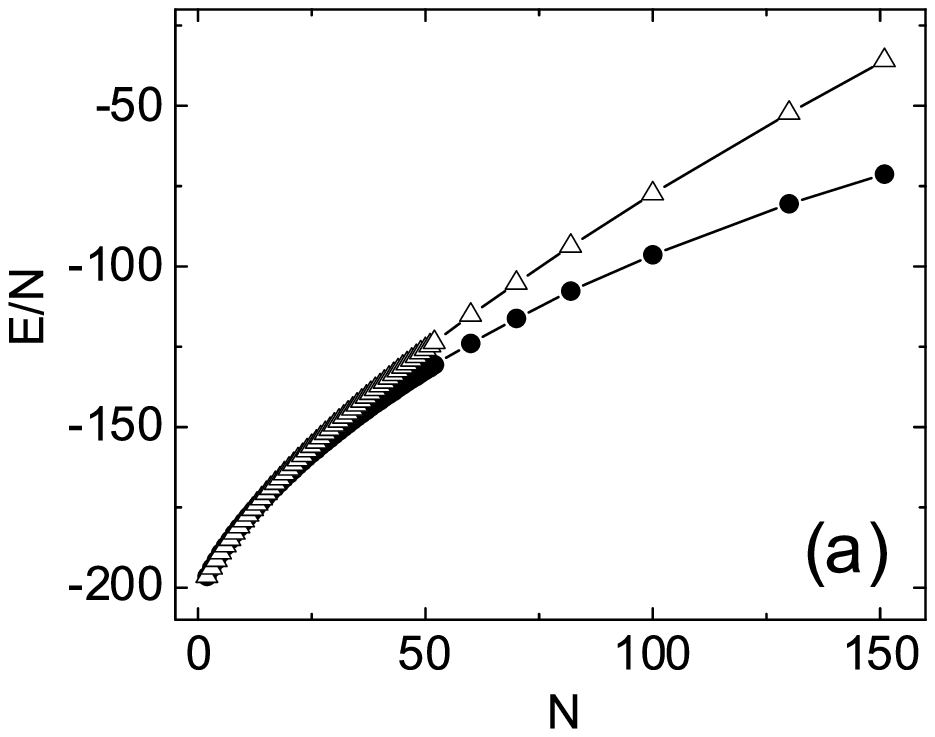}
\end{center}
\vspace{6mm}
\begin{center}
\includegraphics[scale=0.6]{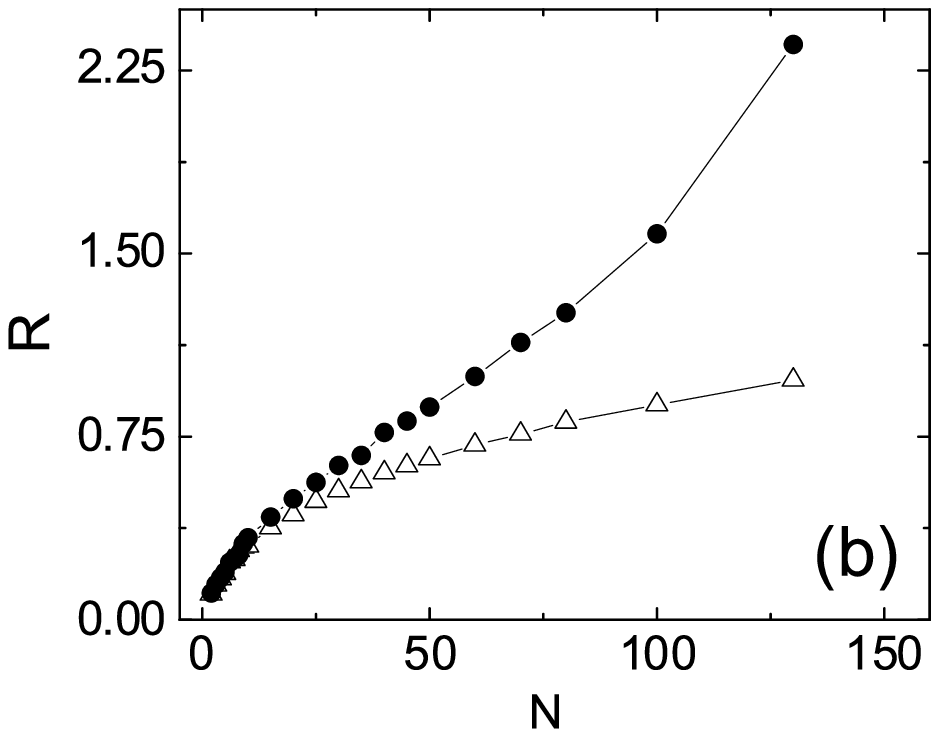}
\caption{\label{fig:fig4}The energy per particle $(a)$ and the
maximum radius $(b)$ of the Coulomb dot (solid circles) and the
parabolic dot (open triangles) as a function of $N$. The
confinement charge of the Coulomb dot is $Z=200$.}
\end{center}
\end{figure}

\begin{figure}\centering
\includegraphics[scale=0.8]{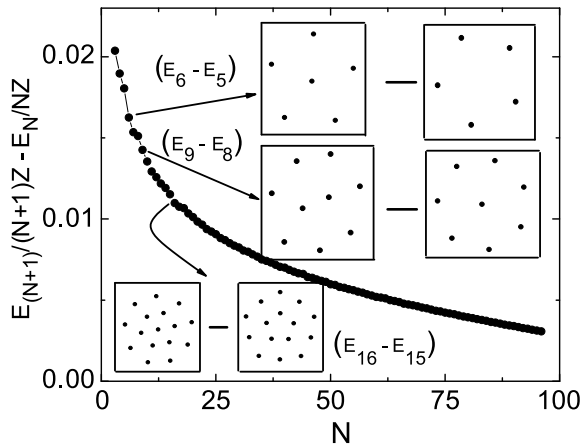}
\caption{Difference in the energy per particle per
$Z$ as a function of $N$ for a Coulomb dot with $Z=100$.
Examples of configurations where kinks are found in the energy curve
are shown in the insets.}
\label{fig:fig5}
\end{figure}

The equivalence of Coulomb screening and the results for a parabolic dot at
small $N$ values is confirmed in Fig.~\ref{fig:fig4}, where the
energy per particle and the maximum radius of a Coulomb dot
confined by a positive impurity charge $Z=200$ are shown as a function of
$N$ and compared with the results obtained for the parabolic dot
\cite{bedanov94,lozovik90}.
More over, if one plots the differences of energy per particle
(scaled by $Z$) as a function of the number of particles $N$, as
is shown in Fig.~\ref{fig:fig5}, we can see small cusps related to
the so called ``magic numbers'' which are known to be an important
feature of the configurations of the parabolic dots \cite{schweg}.

In the opposite case of large electron numbers ($n\lesssim 1$),
where the system is nearly neutral, one can expect to find
features which are specific for our Coulomb screening case and
which are not found with a parabolic dot. In this asymptotic case,
the Coulomb dot presents a low density system \cite{colloid} which
is under the influence of a nonhomogeneous electric field. We
already know that correlation effects in such a system is of most
importance. That is why it is worth to take as a reference the
hydrodynamic approach, which is some mean field theory where no
correlation effects are included.

\section{Hydrodynamic approach}

In the hydrodynamic approach the electrons are described by the 2D
density $\rho({\bf r})$. They create a potential $\phi({\bf r},z)$
which obey the Poisson equation
\begin{equation}\label{poisson}
  \nabla^2\phi(r,z) = 4\pi\rho(r)\delta(z),
\end{equation}
in the whole 3D space. Using the dimensionless variables introduced
in Sec.~\ref{sec:model}, the electron density is measured in
$a^{-2}$ units and the potential in $e/a$ units. Note we took into
account the cylinder symmetry of the problem caused by the
confining potential $V(r) = -Z/\sqrt{r^2+1}$ (see
Eq.~(\ref{reducedhamilt})). The Poisson equation can be replaced
by the Laplace equation
\begin{equation}\label{laplace}
  \nabla^2\phi(r,z) = 0
\end{equation}
everywhere outside the $xy$-plane together with
the boundary condition
\begin{equation}\label{poissonbc}
  \left.\frac{\p\phi(r,z)}{\p z}\right|_{z=0} = 2\pi\rho(r)
\end{equation}
on this plane.

The mathematical model is based on the statement that the electrons
(located in the circular disk of radius $R$) have the same
energy in every point of the dot, namely,
\begin{equation}\label{main}
  \left\{V(r) - \phi(r)\right\}\Big|_{r<R}
  = \mu = \mathrm{const}.
\end{equation}
Here the symbol $\phi({\bf r})=\phi({\bf r},0)$ stands for the
potential created by the 2D electrons in the disk, and the symbol
$\mu$ is the chemical potential of this electron system.

The above equations have to be supplemented with one more
expression, namely, the conservation of the total number of electrons
\begin{equation}\label{number}
  N = 2\pi\int_0^Rrdr\rho(r).
\end{equation}

The hydrodynamic approach is a kind of
mean field theory where no correlation effects are included. For
instance, the phenomena of overcharging \cite{colloid} cannot take
place, and the condition $n = N/Z \le 1$ is always satisfied,
namely, the number of electrons $N$ attracted by the Coulomb
center never exceeds its charge number $Z$. The specific case
$n=1$ is referred to as \textit{complete screening}. In this case
the analytical solution of the above equations can be
obtained using the mirror charge technique, and it leads to the
following result
\begin{equation}\label{fullscreen}
  \rho_{\mathrm{cs}}(r) =
  \displaystyle{\frac{1}{2\pi(r^2+1)^{3/2}}},
\end{equation}
with $n_{\mathrm{cs}} = 1$ and $\mu_{\mathrm{cs}} = 0$.
This simple limiting case result is useful as a reference.

In the general $N<Z$ case  we solved the hydrodynamic
equations using the oblate spherical coordinates
($0<\tau< 1,\,0<\sigma<\infty$) which are defined by
\begin{subequations}\label{oblatesph}
\begin{eqnarray}
  x &=& R\sqrt{(\sigma^2+1)(1-\tau^2)}\cos\theta, \\
  y &=& R\sqrt{(\sigma^2+1)(1-\tau^2)}\sin\theta, \\
\label{oblatesph3}
  z &=& R\sigma\tau
\end{eqnarray}
\end{subequations}
as it was done in Refs.~\cite{ye94,partoens98} for the case of a parabolic
dot with radius $R$. The solution of the Laplace
equation (\ref{laplace}) can be presented as an expansion
\begin{equation}\label{potdisk3D}
  \phi(r,z) = \Phi(\tau,\sigma) = \sum_{n=0}^{\infty}C_nP_{2n}(\tau)
  \frac{Q_{2n}(i\sigma)}{Q_{2n}(0)},
\end{equation}
in terms of the first $P_{2n}$ and second $Q_{2n}$ kind
Legendre polynomials. Thus, the potential created by the electrons
on the disk ($\sigma=0,\;\tau<1$) can be presented as
\begin{equation}\label{potdisk}
  \phi(r) = \Phi(\tau) = \sum_{n=0}^{\infty}C_nP_{2n}(\tau).
\end{equation}
Now taking into account that on the disk we
have according to Eq.~(\ref{oblatesph3})
\begin{equation}\label{deriv}
  \frac{\p}{\p z}\Bigg|_{z=0} =
  \frac{1}{R\tau}\frac{\p}{\p\sigma}\Bigg|_{\sigma=0},
\end{equation}
and satisfying the boundary condition (\ref{poissonbc}) one gets
the analogous expansion for the electron density
\begin{equation}\label{elden}
  \rho(r) = \chi(\tau) =
  -\frac{1}{2\pi R\tau}\sum_{n=0}^{\infty}C_nL_nP_{2n}(\tau),
\end{equation}
where
\begin{equation}\label{notcoeff}
  L_n = -\left\{\frac{d}{d\sigma}\ln Q_{2n}(i\sigma)\right\}
  \Bigg|_{\sigma=0} = 2\left\{\frac{\Gamma(n+1)}{\Gamma(n+1/2)}\right\}^2,
\end{equation}
and $\Gamma(x)$ is the Gamma function.

Next, we expand the Coulomb center potential on the
disk into a series of Legendre polynomials as well
\begin{equation}\label{expcc}
  V(r) = -\frac{Z}{R\sqrt{1+1/R^2-\tau^2}}
  = -\frac{Z}{R}\sum_{n=0}^{\infty}A_nP_{2n}(\tau),
\end{equation}
and inserting it together with Eq.~(\ref{potdisk}) into
Eq.~(\ref{main}) we obtain the final expression
\begin{equation}\label{solpot}
  C_n = -\mu\delta_{n,0} - \frac{Z}{R}A_n
\end{equation}
for the electron density expansion coefficients $C_n$.

In order to define the chemical potential $\mu$ we have to
remember that the electron density has to be equal to zero at the
free electron system boundary $r=R$. Thus, the electron density
should satisfy the following condition
\begin{equation}\label{fermi}
\begin{split}
  2\pi R\lim_{\tau\to 0}\tau\chi(\tau)
  &= -\sum_{n=0}^{\infty}C_nL_nP_{2n}(0) \\
  &= -\mu L_0 - \frac{Z}{R}\sum_{n=0}^{\infty}A_nL_nP_{2n}(0) = 0,
\end{split}
\end{equation}
what finally enables us to define the chemical potential
\begin{eqnarray}\label{chempot}
  \mu &=& -\frac{Z}{RL_0}\sum_{n=0}^{\infty}A_nL_nP_{2n}(0) \nonumber \\
  &=& -\frac{Z\sqrt{\pi}}{R}\sum_{n=0}^{\infty}
  (-1)^nA_n\frac{\Gamma(n+1)}{\Gamma(n+1/2)},
\end{eqnarray}
and write down the following electron density expression
\begin{equation}\label{eldensfin}
  \chi(\tau) =
  \frac{Z}{2\pi R^2\tau}\sum_{n=1}^{\infty}A_nL_n\left\{P_{2n}(\tau)
  -P_{2n}(0)\right\}.
\end{equation}
Inserting the density expression (\ref{elden}) into
Eq.~(\ref{number}) one obtains the number of electrons
\begin{equation}\label{nfin}
  n = \frac{N}{Z} = \frac{2}{\pi}\left(A_0+\frac{R}{Z}\mu\right),
\end{equation}
taking part in the screening of the Coulomb center.

Equations (\ref{expcc}), (\ref{chempot}), (\ref{eldensfin}), and
(\ref{nfin}) actually are the solution of the problem if the
coefficients $A_n$ are known. These coefficients are calculated in
Appendix~\ref{app1}.

The density (\ref{eldensfin}) enables us to calculate all
properties of the electron system. For instance, the square of
mean electron radius can be estimated as
\begin{eqnarray}\label{mrad}
  <r^2> = \frac{1}{N}\int_0^Rdr\,r^3\rho(\tau)
  = R^2\left(\frac{2}{3}-\frac{16A_1}{15\pi n}\right).
\end{eqnarray}

\begin{figure}\centering
\includegraphics[scale=0.35]{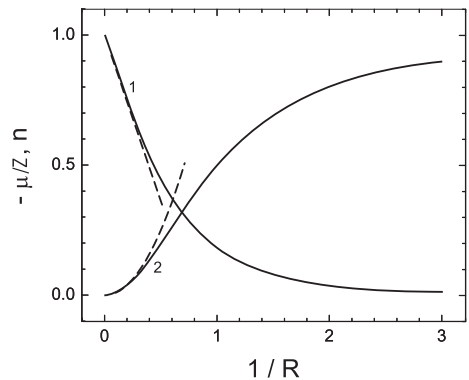}
\caption{Hydrodynamic solution (solid curves) and its asymptotic
(dashed curves) for large $R$: 1 -- number of particles
(Eq.~(\ref{nfin})) , 2 -- chemical potential
(Eq.~(\ref{chempot})).} \label{fig:fig6} \vspace{3mm}
\end{figure}

\begin{figure}\centering
\includegraphics[scale=0.3]{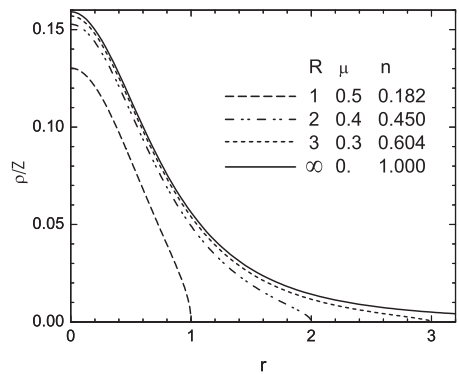}
\caption{Hydrodynamic solution for the electron density for
various dot radii.} \label{fig:fig7}
\end{figure}

Due to the linear dependence of the confinement potential
(\ref{expcc}), the chemical potential (\ref{chempot}) and the
density (\ref{eldensfin}) on the charge number $Z$, it can be
removed by a scaling, and the hydrodynamic problem is actually
controlled by a single parameter (say, $n$ or $R$) in contrast to
the system of discrete electrons for which the two parameters ($Z$
and $N$) were essential. Thus, the general solution can be
presented by both chemical potential and density curves, as is
shown in Fig.~\ref{fig:fig6} by solid curves.
The corresponding dashed curves indicate the asymptotic
of the above quantities
\begin{subequations}\label{asymptmun}
\begin{eqnarray}
\label{asymptmun1}
  \mu_{\mathrm{as}} &=& -\frac{Z}{R^2}, \\
\label{asymptmun2}
  n_{\mathrm{as}} &=& 1 - \frac{4}{\pi R},
\end{eqnarray}
\end{subequations}
which holds for a nearly neutral ($n\to 1$) electron system. In
Fig.~\ref{fig:fig7} the electron density for various dot radii is
shown.
We see that in the case of small radius (or equivalently large
$Z$, i.e. $n<<1$) the electron density becomes similar to the
density in a parabolic dot ($\rho\sim\sqrt{1-r^2/R^2}$), while in
the opposite large radius case it tends to the limiting density
(\ref{fullscreen}) for the completely screened case.

\section{Correlation effects}

\begin{figure}
\begin{center}
\includegraphics[scale=0.33]{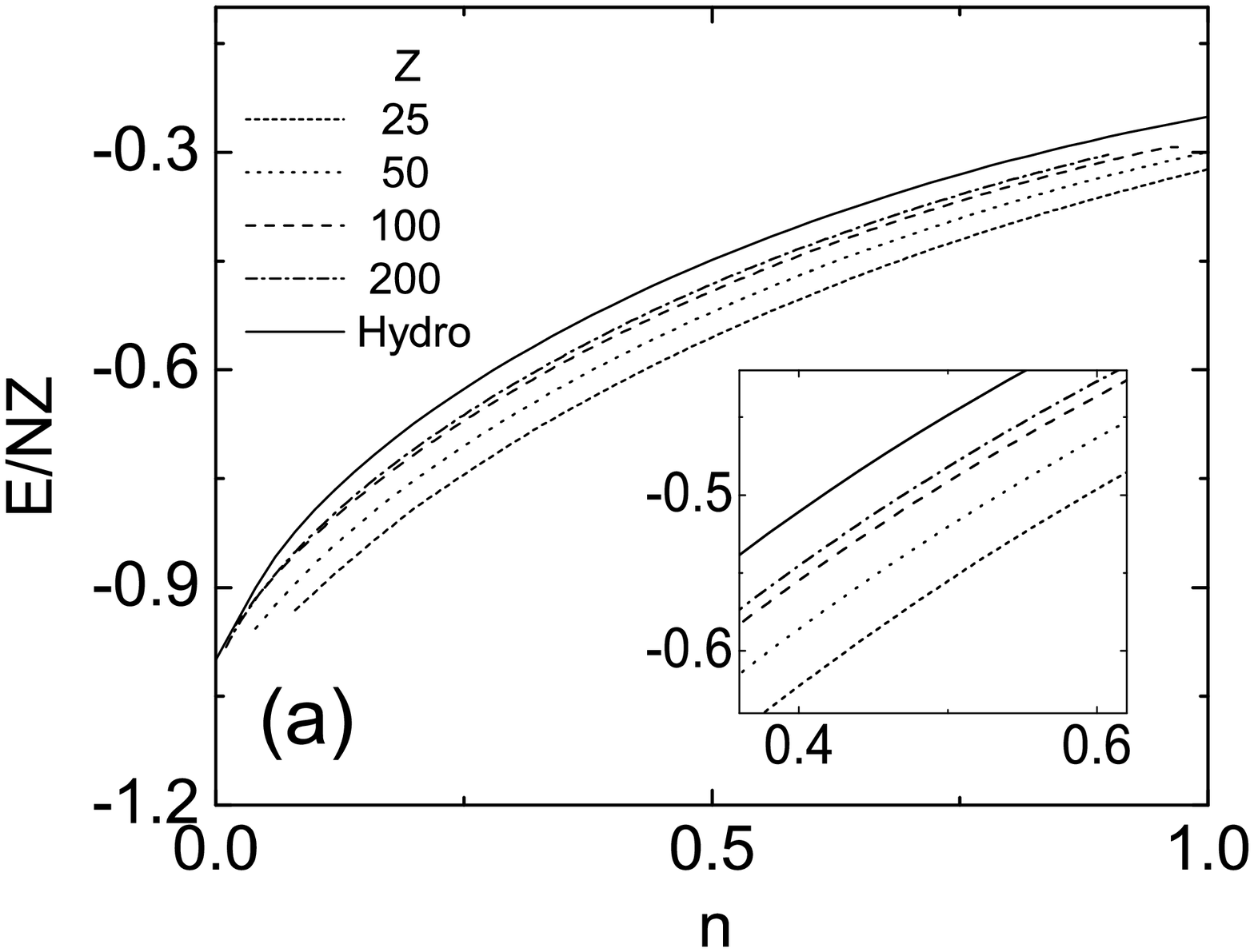}
\end{center}
\vspace{6mm}
\begin{center}
\includegraphics[scale=0.33]{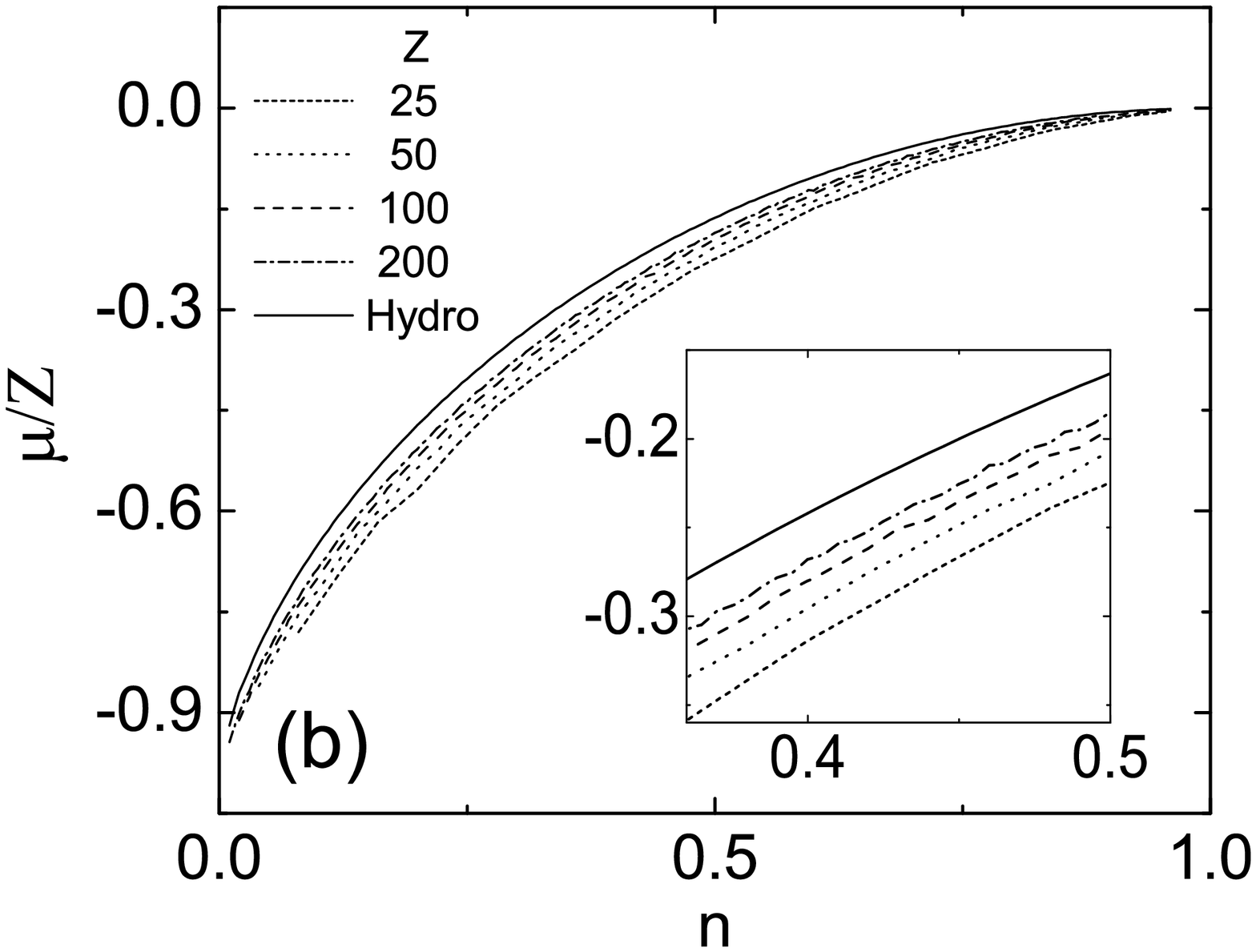}
\caption{The scaled energy per particle (a) and chemical potential
(b) obtained from our numerical simulations for different values
of the confinement charge $Z$ are compared with the hydrodynamic
result.} \label{fig:fig8}
\end{center}
\end{figure}

In order to compare the numerical simulation results with the
hydrodynamic approach we scaled them by the charge number $Z$. In
Fig.~\ref{fig:fig8}(a), the energy per particle scaled by $Z$ is
shown as a function of the relative number of electrons $n=N/Z$.

The hydrodynamic result (solid curve) was obtained by numerically
integrating the scaled chemical potential (\ref{chempot}) over the
number of electrons $n$, namely,
\begin{equation}\label{hydroenergy}
  \frac{E(n)}{ZN} = \frac{1}{ZN}\int_0^n\mu dn.
\end{equation}
In Fig.~\ref{fig:fig8}(b) the same comparison is shown for the
scaled chemical potential. The numerical simulation results were
calculated as the difference of the total energies for adjacent
configurations, namely, $\mu_N = E_{N+1}-E_N$. We see that when
the charge number $Z$ increases the curves tend towards the
hydrodynamic result. This is more clearly seen for the chemical
potential (Fig.~\ref{fig:fig8}(b)) than for the total energy
curves (Fig.~\ref{fig:fig8}(a)).

\begin{figure}
\begin{center}
\includegraphics[scale=0.65]{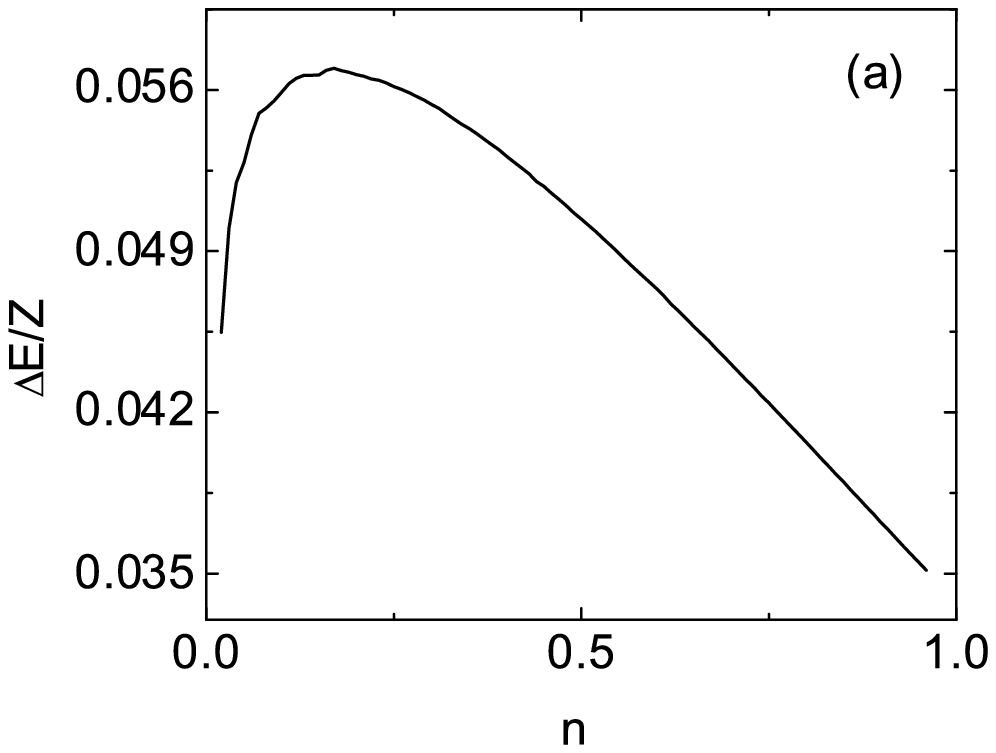}
\end{center}
\vspace{6mm}
\begin{center}
\includegraphics[scale=0.65]{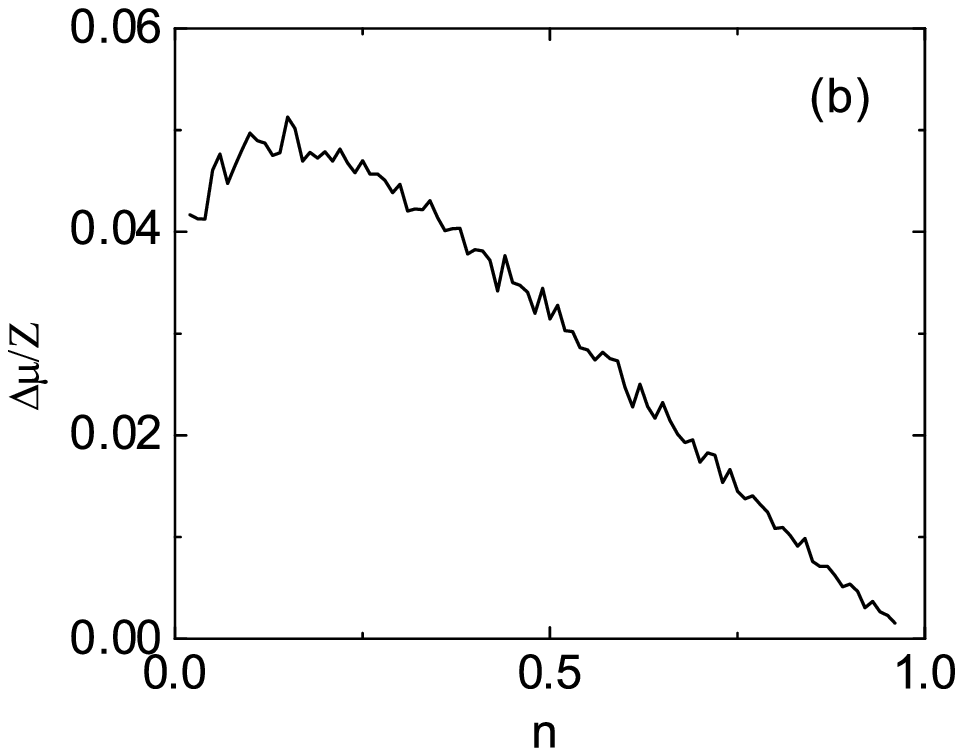}
\caption{The difference in energy (a) and chemical potential (b)
between the hydrodynamic result and the simulation result
for $Z=100$.}
\label{fig:fig9}
\end{center}
\end{figure}

The deviation of the energy and the chemical potential of the
'exact' numerical simulation results from their hydrodynamic
counterparts has to be interpreted as a \textit{correlation
energy}, because such correlations are not included in the
hydrodynamic approach. As an example, we show in
Fig.~\ref{fig:fig9}, for $Z=100$,  this difference in energy per
particle and similar results for the chemical potential. The
curves for other values of the confinement charge demonstrate a
similar behavior.
The most remarkable feature of these curves is the linear behavior
in the nearly full screening region ($n\to 1$). The
curves in the above region can be fitted by the rather simple
analytical expressions:
\begin{subequations}\label{fit}
\begin{eqnarray}
  \Delta E/NZ = A_E-B_En, \\
  \Delta \mu/Z = A_{\mu}-B_{\mu}n.
\end{eqnarray}
\end{subequations}
For the results of Fig.~\ref{fig:fig9} we found
\begin{subequations}\label{examplefit}
\begin{eqnarray}
  \Delta E/NZ = 0.068 - 0.034n, \\
  \Delta \mu/Z = 0.065-0.067n.
\end{eqnarray}
\end{subequations}
Although the coefficients of these expressions are $Z$-dependent,
we found that
the ratio of them are Z-independent ($A_E/B_E\approx 2$ and
$A_{\mu}/B_{\mu}\approx 1$).
\begin{figure} \centering
\includegraphics[scale=0.65]{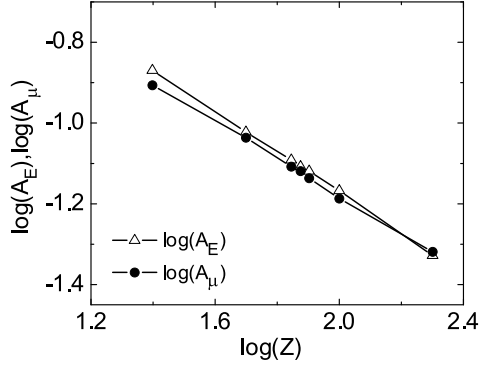}
\caption{Logarithmic plot of the parameter $A$ of the linear
fit of the deviation curves for the energy per particle and the
chemical potential in the nearly \textit{complete screening} limit.}
\label{fig:fig10}
\end{figure}
The linear behavior of these curves in the double logarithmic
plot, as it is seen in Fig.~\ref{fig:fig10}, enables us to fit
them by $A=\alpha Z^{-\beta}$. The straight lines in
Fig.~\ref{fig:fig10} correspond to $\alpha=0.69$ and $\beta=0.5$
for $A_E$, and $\alpha=0.55$ and $\beta=0.46$ for $A_{\mu}$. Therefore,
we suggest the following asymptotic behaviour for $N \rightarrow Z$
\begin{subequations}\label{fulfit}
\begin{eqnarray}
\label{fulfit1}
  \Delta E/N \sim \sqrt{Z}\left(1-\frac{N}{2Z}\right), \\
\label{fulfit2}
  \Delta\mu \sim \sqrt{Z}\left(1-\frac{N}{Z}\right).
\end{eqnarray}
\end{subequations}
These results express the contribution of
correlation to the energy and
the chemical potential of our Coulomb bound classical dot.

Qualitatively, such dependence can be explained
by means of the crystallization energy which can be estimated
as
\begin{equation}\label{cren}
  E_{\mathrm{cr}} = \int_0^Rd^2r\rho(r)E_I(r)
\end{equation}
where $E_I(r)$ is the local density crystallization energy
which we approximated by the result from a homogeneous Wigner crystal
\cite{bonsal77}
\begin{equation}\label{bonsal}
  E_I = a_0\sqrt{\rho}.
\end{equation}
The coefficient equals $a_0\approx 3.921$ for the triangular
lattice. Now inserting the above expression into Eq.~(\ref{cren}),
replacing the electron density by its asymptotic expression
(\ref{fullscreen}), and taking Eq.~(\ref{asymptmun2}) into account
we obtained the following estimate:
\begin{eqnarray}\label{estim}
  E_{\mathrm{cr}} &\approx& 2\pi a_0\int_0^Rrdr\rho_{\mathrm{cs}}^{3/2}
  = \frac{a_0Z^{3/2}}{\sqrt{2\pi}}\int_0^R\frac{rdr}{(r^2+1)^{9/4}} \nonumber \\
  &=& \frac{2a_0Z^{3/2}}{5\sqrt{2\pi}}\left\{1-\left[\frac{\pi}{4}
  (1-n)\right]^{5/2}\right\}.
\end{eqnarray}
Taking only the first term into account and dividing the above
crystallization energy by $N=Z\{1-(1-n)\}$ we obtained the following
approximate crystallization energy per particle
\begin{equation}\label{crenp}
  \frac{E_{cr}}{N} \approx \frac{2a_0\sqrt{Z}}{5\sqrt{2\pi}}(2-n),
\end{equation}
which demonstrates the same parameter dependencies as obtained
earlier, Eq. (\ref{fulfit1}), from a fitting of our simulation
results.

Differentiating the above crystallization energy expression by $n$
we also obtained an estimation for the correlation contribution to
the chemical potential:
\begin{eqnarray}\label{crchempot}
  \mu_{cr} = \frac{d}{dN}E_{cr}
  = \frac{2a_0\sqrt{Z}}{5\sqrt{2\pi}}\frac{d}{dn}n(2-n)
  \sim \sqrt{Z}(1-n)
\end{eqnarray}
which coincides with our earlier result, Eq. (\ref{fulfit2}).

\begin{figure}
\begin{center}
\includegraphics[scale=0.37]{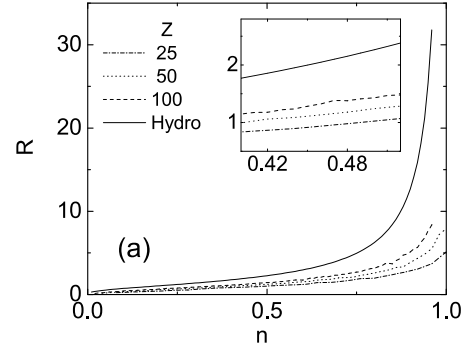}
\end{center}
\vspace{6mm}
\begin{center}
\includegraphics[scale=0.36]{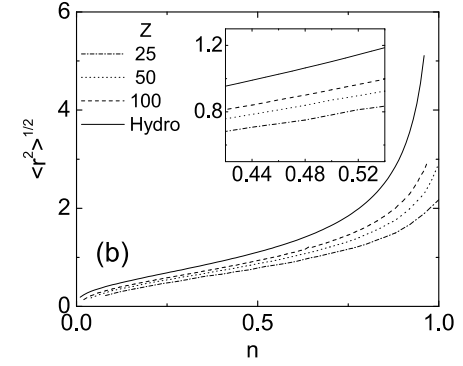}
\caption{Comparison of maximal ($R$) and mean radii ($r$) of the
electron system obtained from the numerical simulation (for
different $Z$-values) and in the hydrodynamic approach as function
of $n=N/z$. The
Z-dependence of the maximum and the mean radius is enlarged in the
inset.} \label{fig:fig11}
\end{center}
\end{figure}

Unfortunately, no similar simple expressions can be obtained for
the dot radius as depicted in Fig.~\ref{fig:fig11}. We see that
the radii from the hydrodynamic approach and the discrete system
differs substantially, and the deviation grows in the limiting
$n\to 1$ case. The discrete system is more compact as compared
with the continuous one, which is an indication that the discrete
system may lead to overcharging \cite{colloid} which cannot be
described by a hydrodynamic theory.

\section{Conclusions}

We studied numerically the ground state properties of a
2D model system consisting of classical charged particles
which are Coulomb bound. This system is similar to an atomic
system. The $N$ electrons moving in a plane are confined through a
positive remote impurity potential of charge $Z$. The configurations for
$n \ll N/Z$ are similar to those of a parabolic confined dot, but
for $n\sim 1$ they are very different, with an inner core consisting of
approximately a triangular lattice and an outer region with
particles situated on a ring.

A hydrodynamic analysis of the problem clearly emphasizes the
importance of correlation effects between the negatively charges
particles. Furthermore, analytical expressions for the correlation
energy contributions to the total energy and the chemical
potential were obtained in the limit of nearly overscreening,
e.~g.~$n\sim 1$. Those analytical results compare favorably well
with our numerical ``exact'' simulation results.

\acknowledgements

W.~P.~Ferreira and G.~A.~Farias are supported by the
Brazilian National Research Council(CNPq) and the
Ministry of Planning (FINEP). Part of this work was supported by
the Flemish Science Foundation (FWO-Vl), the ``Onderzoeksraad van de
Universiteit Antwerpen" (GOA) and the EU-RTN on ``Surface electrons".

\appendix

\section{Coulomb potential expansion coefficients}
\label{app1}

According to Eq.~(\ref{expcc}) the Coulomb potential expansion
coefficients $A_n$ depend on the single variable $b=1/R$, and they
can be calculated straightforwardly. Indeed, multiplying
Eq.~(\ref{expcc}) by $P_{2m}(\tau)$, integrating it over $\tau$
and using the normalization integral for symmetric Legendre
polynomials one obtains the following integral expression:
\begin{equation}\label{integcoeff}
  A_{n}(b) = (4n+1)\int_0^1
  \frac{P_{2n}(\tau)d\tau}{\sqrt{1+b^2-\tau^2}}.
\end{equation}
The most accurate way to calculate the above integral is to expand
the denominator into a $\tau^{2n}$ power series and use the
analytical expression for
\begin{equation}\label{ryzhik}
  \int_0^1x^{2k}P_{2n}(x)dx
  = \frac{(-1)^n\Gamma(n-k)\Gamma(k+1/2)}
  {2\Gamma(-k)\Gamma(n+k+3/2)},
\end{equation}
what enables us to convert the integral (\ref{integcoeff}) into the
following sum
\begin{eqnarray}\label{coeffa}
  A_{n}(b) &=& \frac{(4n+1)}{2\sqrt{\pi}\,(1+b^2)^{n+1/2}} \nonumber \\
  && \times\sum_{m=0}^{\infty}
  \frac{\Gamma^2(m+n+1/2)(1+b^2)^{-m}}{\Gamma(m+1)\Gamma(m+2n+3/2)}.
\end{eqnarray}
The convergence of this sum is rather slow. Fortunately, it can be
improved by adding the asymptotic, namely, replacing
Eq.~(\ref{coeffa}) by the following expression
\begin{eqnarray}\label{coeffas}
  A_{n}(b) &=& \frac{(4n+1)}{2\sqrt{\pi}\,(1+b^2)^{n+1/2}} \nonumber \\
  &\times& \sum_{m=0}^{s}
  \frac{\Gamma^2(m+n+1/2)(1+b^2)^{-m}}{\Gamma(m+1)\Gamma(m+2n+3/2)} \nonumber \\
  &+& \frac{4n+1}{(1+b^2)^{s+n+1/2}} \nonumber \\
  &\times& \left\{\frac{1}{\sqrt{\pi s}}- b\,e^{b^2s}
  \left[1-\Phi(b\sqrt{s})\right]\right\}.\phantom{m}
\end{eqnarray}
where $s$ is an integer which should be optimized for rapid
convergence and $\Phi(z)$ is the error function. The above expression
is very convenient and enables us to obtain good accuracy in a
fast way if one uses the recurrence expressions for the
calculations of the gamma functions (as for the Legendre
polynomials in Eq.~(\ref{eldensfin}) as well).


\begin{thebibliography}{99}
\bibitem{jacak98}L.~Jacak, P.~Hawrylak, and A.~W\'{o}js,
\textit{Quantum dots} (Springer-Verlag, Berlin, 1998).
\bibitem{bedanov94}V.~M.~Bedanov and F.~M.~Peeters, Phys.~Rev.~B {\bf49},
2667 (1994).
\bibitem{gil96}G.~A.~Farias and F.~M.~Peeters, Solid~State~Commun. {\bf100},
711 (1996).
\bibitem{watanabe86}H.~Watanabe and T.Inoshita, Optoelectron.\ Dev.\
Technol.\ \textbf{1}, 33 (1986).
\bibitem{inshita88}T.~Inoshita, S.~Ohnishi, and A.~Oshiyama,
Phys.\ Rev.\ B \textbf{38}, 3733 (1988);
E.~A.~Andryushin and A.~P.~Silin, Sov.\ Phys.\ Solid State
\textbf{33}, 123 (1991) [Fiz.\ Tverd.\ Tela (Leningrad)
\textbf{33}, 211 (1991)].
\bibitem{yannou}See e.g., C.~Yannouleas and U.~Landman,
cond-mat/0204530; A.~Matulis and F.~M.~Peeters, Solid~State~Commun.
{\bf117}, 655 (2001).
\bibitem{colloid}See e.g. B.~I.~Shklovskii, Phys.~Rev.~Lett. {\bf82},
3268 (1999); R.~Messina, C.~Holm, and K.~Kremer, Phys.~Rev.~E {\bf
64}, 21405 (2001).
\bibitem{andrei}See e.g. {\it Two-dimensional electron systems on
helium and other substrates}, edited by E.~Y.~Andrei (Dordrecht,
Kluwer, 1997).
\bibitem{marmorkos} I.K. Marmorkos, V.A. Schweigert, and F.M. Peeters,
Phys. Rev. B {\bf 55}, 5065 (1997); C. Riva, V.A. Schweigert, and F.M. Peeters,
Phys. Rev. B {\bf 57}, 15392 (1998)..
\bibitem{metropolis}N.~Metropolis, A.~W.~Rosenbluth, M.~N.~Rosenbluth,
A.~M.~Teller, and E.~Teller, J.~Chem.~Phys, {\bf21}, 1087 (1953).
\bibitem{ye94}Z.~L.~Ye and E.~Zaremba, Phys.~Rev.~B {\bf50},
17217 (1994).
\bibitem{schweg}V.~A.~Schweigert and F.~M.~Peeters, Phys.~Rev.~B {\bf51},
7700 (1995).
\bibitem{partoens98}B.~Partoens, A.~Matulis, and F.~M.~Peeters,
Phys.~Rev.~B \textbf{57}, 13039 (1998).
\bibitem{lozovik90}Y.~E.~Lozovik and V.~A.~Pomirchy, Phys.~Status.~Solidi B
{\bf161}, K11 (1990).
\bibitem{bonsal77}L.~Bonsall and A.~A.~Maradudin, Phys.~Rev B
{\bf 15}, 1959 (1977).
\end{thebibliography}
\end{document}